\documentclass[%
 reprint,
 amsmath,amssymb,
 aps,
pra,
]{revtex4}

\usepackage[english]{babel}
\usepackage[T1]{fontenc}

\usepackage{graphicx}% Include figure files
\usepackage{dcolumn}% Align table columns on decimal point
\usepackage{bm}% bold math
\begin{document}

\preprint{APS/123-QED}

\title{An Analytic Study of the Wiedemann-Franz Law and the Thermoelectric Figure of Merit}% Force line breaks with \\

\author{Aakash Yadav}
\author{PC Deshmukh}%
\affiliation{%
 Indian Institute of Technology Tirupati, Tirupati 517506, India.
}%

\author{Ken Roberts}
\author{NM Jisrawi}
\altaffiliation[Also at the ]{Centre for Advanced Materials Research, Sharjah, United Arab Emirates.}%
\author{SR Valluri}
\altaffiliation[Also at the ]{ Dept of Management, Economics, and Mathematics, Kings University College, University of Western Ontario.}%
\email[Corresponding author: ]{valluri@uwo.ca}
\affiliation{Department of Physics and Astronomy, University of Western Ontario, London,ON N6A 3K7, Canada.}%

\date{\today}% It is always \today, today,

\begin{abstract}
 Advances in optimizing thermoelectric material efficiency have seen a parallel activity in theoretical and computational advances. In the current work, it is shown that the calculation of exact Fermi-Dirac integrals enables the generalization of the Wiedemann-Franz law (WF) to optimize the dimensionless thermoelectric figure of merit $ ZT = {{\alpha ^2 \sigma } / {\kappa}} $. This is done by optimizing the Seebeck coefficient $\alpha$, the electrical conductivity $ \sigma $ and the thermal conductivity $\kappa $. In the calculation of the thermal conductivity $\kappa$, both electronic and phononic contributions are included. The solutions provide insight into the relevant parameter space including the physical significance of complex solutions and their dependance on the scattering parameter $r$ and the reduced chemical potential $\mu^*$.
\end{abstract}

\maketitle

%\tableofcontents

\section{Introduction}

Though a 19th century science, thermoelectricity has seen a revival of late due to the perceived dangers of fossil fuels and global warming. Thermoelectric (TE) materials continue to generate interest despite the fact that the phenomena have been known since the 19th century \cite{Seebeck1822,Peltier1834,Thomson_1857}. The renewed interest has been stimulated by concern over the effects of fossil fuels as major contributors to global warming as well as the advancement in materials properties for energy generation and conversion. Thermoelectric materials can generate power using the Seebeck effect or refrigerate using the Peltier effect. They are capable of converting heat flow directly into electrical energy or vice-versa\cite{Goldsmid2009}. Efforts are focused on the enhancement of the material efficiencies by optimizing the TE figure of merit. A recent review by Morelli has identified three material families that are the focus of $ZT$ enhancement research: Phonon-glass-electron crystal (PGEC), Bulk nano-structured materials, and Crystals with large anharmonicity \cite{Morelli2017}. The PGEC class of materials use the concept of minimum thermal conductivity via a phonon-type mechanism to achieve a figure of merit as high as 1.5 in n-type \textit{skutterudites}\cite{Morelli_1995}. In the early 1990s, Hicks and Dresselhaus \cite{Hsu_2004, Fleischmann_1961, Irie_1963} studied the 1-D and 2-D confinement leading to a positive effect for nano-structured materials generally known as LAST (Lead-antimony-silver-telluride) materials. Figures of merit as high as 2.2 were reached in doped samples. The third class of materials which has seen intense investigations of late are based on lowering thermal conductivity in materials with large anharmonicity in the crystal structure and a large Grueneissen coefficient $\gamma$\cite{Ashcroft1976}. A value of $\gamma = 0$  describes the harmonic lattice with a large thermal conductivity. Tetrahedrites are such a family of compounds that contain so-called lone electron pairs which enhance anharmonicity with distinctly coordinated Cu atoms. This leads to lower thermal conductivity and a $ZT$ of about 0.9 \cite{Lu_2012, Lu_2013, Lu2013, Heo2014, liu-zhou}.  

In addition to intensive activity in enhancing material properties, theoretical understanding has kept pace and resulted in a widely accepted formalism based on Fermi-Dirac statistics that computes the figure of merit in terms of Fermi-Dirac integrals and relates efficiency to the TE figure of merit $ZT$\cite{Dingle1957,zhao_2014}. The polylogarithm functions \cite{murali2011}, the {LambertW} function \cite{cpj2000} and its generalization have created a renaissance in the solution of diverse problems that include applications to thermoelectric materials.

The current work provides a generalization of the Wiedemann-Franz law (WF) which plays an important role in thermoelectric material research\cite{Kittel2004}. In section \ref{WF-law}, we derive  corrections to the WF law based on the widely-used Fermi-Dirac integrals describing the thermoelectric properties of semiconducting materials \cite{Goldsmid2009} and demonstrate how the corrections vary with the scattering parameter, $r$ and the reduced chemical potential, $\mu^* = E_f/kT$. In section \ref{TC-extema},  we extremize the expression for thermal conductivity (in the form of the polylogarithmic functions)  with respect to reduced chemical potential, $\mu^*$ and temperature $T$, respectively. In section \ref{EC-extrema}, we provide insight into the mathematical expressions resulting from the extremization of the electrical conductivity with respect to temperature and relate that to maximizing the figure of merit. In section \ref{lat-extrema}, we reexamine the phonon contribution to thermal conductivity and in section \ref{conclusions}, we present our conclusions.

\section{The Wiedemann-Franz Law} \label{WF-law}

The Wiedemann-Franz Law (1853) states that the ratio of the electronic contribution of the thermal conductivity $\kappa$ to the electrical conductivity $\sigma$ of a metal is proportional to the temperature $ T $ based on a semi-classical treatment of the electron gas\cite{Ashcroft1976, Kittel2004}. 

\begin{align}
\frac{\kappa}{\sigma} =L_0T
\end{align}
where $L_0$ is a constant called the Lorenz number is given by:
\begin{align}
L_0 = \left(\frac{k}{e}\right)^2\frac{\pi^2}{3}
\end{align}
\begin{align}
L_0 = 2.44\times10^{-8}W\Omega K^{-2}     \notag
\end{align}
Here $k$ is the Boltzmann constant and $e$ is the electronic charge.
For over 150 years, the Wiedemann-Franz law has proven to be roughly stable amongst the multitude of metallic systems that have been studied\cite{kaye+laby}.
But recent experiments over a couple of decades show that there are several limitations to the law, the value of Lorenz number L is not the same for every material and the law does not hold for intermediate temperatures. Experiments have shown that the value of Lorenz number, $L$, while roughly constant, is not exactly the same for all materials\cite{Wakeham_2011}. In the realm of space physics, Bespalov and Savina \cite{MNRAS2007} have shown that the turbulent plasma conductivity along with the anomalous thermal conductivity of the medium result in the generalization of the $WF$ law. 
In many high purity metals, both the electrical and thermal conductivities rise as the temperature is decreased. In certain materials (such as silver or aluminum), however, the value of $L_0$  may also decrease with temperature. In the purest samples of silver and at very low temperatures, L can drop by as much as a factor of 10\cite{MNRAS2007}. 

The standard treatment of the generalized WF law is based on the Fermi Dirac distribution of electrons and holes in semiconductor materials\cite{Goldsmid2009}. Plugging in the equations for $\kappa$ and $\sigma$ in terms of Fermi-Dirac integrals as shown in equation \ref{FermiDirac} below, it becomes apparent that the Lorenz number $L_0$ is not a constant. Instead, it converges to $ (\frac{k}{e})^2 \frac{\pi^2}{3} $ for higher values of the reduced chemical potential $\mu^*$, irrespective of the value of the scattering parameter $r$:

\begin{align}
\sigma = \frac{16\pi m e^2 l_0 \left(kT\right)^{r+1}F_r}{3h^3}
\end{align}

\begin{align}
\kappa_e = \frac{16\pi m l_0 k^{r+3} T^{r+2}}{3h^3}\left[\left(r+3\right)F_{r+2} - \frac{\left(r+2\right)^2 F^2 _{r+1}}{\left(r+1\right)F_r}\right]
\end{align}

Use of the relationship connecting the Fermi integral $F_r$ to the Polylogarithm functions $Li_{r+1}(z)$ \cite{NIST-math} where $z = -e^{\mu^*}$, we get, 

\begin{align} \label{FermiDirac}
F_r\left(\mu^*\right)=-\Gamma\left(r+1\right)Li_{r+1}\left(-e^{\mu^*}\right)
\end{align}

This result leads to a generalized expression for the Lorenz number in the form:

\begin{align}
L = \left(\frac{k}{e}\right)^2A\left(r,\mu^*\right)
\end{align}

where the function $A\left(r,\mu^*\right)$ can be expressed more concisely in terms of polylogarithms.

\begin{align}
A\left(r,\mu^*\right)=\frac{\left(r+2\right)\left(r+3\right)Li_{r+1}Li_{r+3}-\left(r+2\right)^2Li_{r+2}^2}{Li_{r+1}^2}
\end{align}

where the argument $z$ in the Polylog functions is henceforth omitted.

\begin{figure}[h!]
\centering
\includegraphics[scale=.3]{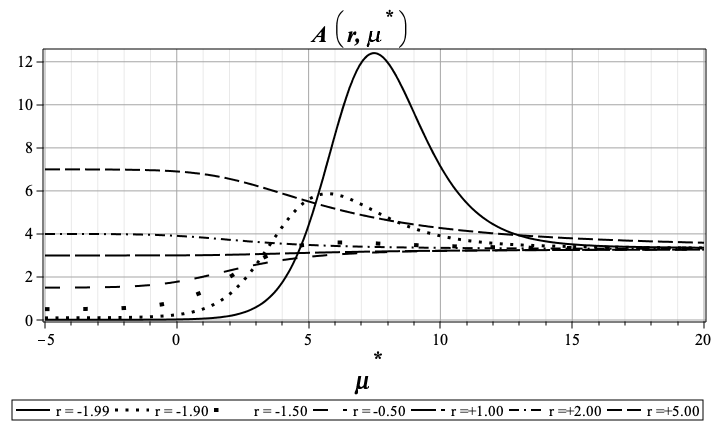}
\caption{Scaling of the WF law by $A\left(r,\mu^*\right)$}
\label{fig:wf_law}
\end{figure}

 The function $A(r,\mu ^*$) is plotted in figure 1 against $\mu^*$ for several indices. The plot shows that the expression converges to the semi-classical value of the WF law for the Lorenz number. The expression for $A(r,\mu^*)$ converges for a wide range of $\mu^*$ as illustrated in the figure for values of $r > -2.0$. For positive values of $\mu^*$, evaluation of the first and second derivatives have shown that there are no local maxima but there are inflection points signifying critical values in the range $0 < r < 5.0$. For negative values of $r$,  $A(r,\mu ^*$) exhibits maxima. The important question of whether materials with negative $r$ exist and the ranges over which they do will be discussed in a future study.

Fermi-Dirac integrals play an important role in the study of semiconductors and appear frequently in the treatment of thermal properties. The corrections to the WF law are embodied in the detailed expressions of the Fermi-Dirac integrals derived below in terms of polylogs.

The Fermi Dirac integral of index $n$ is defined as:

\begin{align}
F_n\left(\mu^*\right)=\int_{0}^{\infty}\frac{ x^n}{e^{x-\mu^*}+1}dx
\end{align}

where $\mu^*$ is the reduced chemical potential and $x$ is the electron energy in units of $k_B T$. 
Computer algebras like Mathematica can compute this expression in closed form for $(Re(n) > -1 )$ in terms of polylogs as 
$ F_n(\mu^*)= -\Gamma \left( n+1 \right) Li_{n+1} ( -{e}^{\mu})$.
The physical details of the restrictions on the values of $r$ and $\mu^*$, however, make the actual detailed calculations of the integrals included in the expression for $A(r,\mu^*)$ a useful exercise. We will leave the detailed calculation to the appendices but will note the outline of the calculation here. 
We start by splitting the integral into two parts separated by some physical value of $\mu^*$. Simplification of (8) above gives:

\begin{align*}
 \int_{0}^{\infty}\frac{ x^n}{e^{x-\mu^*}+1}dx =
\end{align*}
\begin{align}
\int_{0}^{\mu^*}\frac{ x^n \left(1+e^{x-\mu^*}-e^{x-\mu^*}\right)}{e^{x-\mu^*}+1}dx+\int_{\mu^*}^{\infty}\frac{ x^n}{e^{x-\mu^*}+1}dx
\end{align}

The first term above can be written as:

\begin{align*}
    \int_{0}^{\mu^*}\frac{ x^n \left(1+e^{x-\mu^*}-e^{x-\mu^*}\right)}{e^{x-\mu^*}+1}dx=
\end{align*}
\begin{align}
\int_{0}^{\mu^*}x^ndx-\int_{0}^{\mu^*}\frac{ x^n e^{x-\mu^*}}{e^{x-\mu^*}+1}dx
\end{align}

Using the transformation $x-\mu^*=z$, 

\begin{align*}
    \int_{0}^{\mu^*}\frac{ x^n e^{x-\mu^*}}{e^{x-\mu^*}+1}dx = 
\end{align*}
\begin{align}
- \int_{-\mu^*}^{0}\frac{ e^z \left(\mu^*+z\right)^n}{e^{z}+1}dz = - \int_{-\mu^*}^{0}\frac{ e^z\mu^{*^n} \left(1+\frac{z}{\mu^*}\right)^n}{e^{z}+1}dz
\end{align}

Using the binomial expansion for $z/\mu^*  < 1 $, we get

\begin{align}
 = - \int_{-\mu^*}^{0}\frac{ e^z\mu^{*^n} \left(1+\frac{nz}{\mu^*}+\frac{n\left(n-1\right)}{2}\frac{z^2}{\mu^{*^2}}...\right)}{e^{z}+1}dz
\end{align}

\begin{align*}
     = - \bigg[ \mu^{*^n}ln\left(1+e^z\right)\Biggr|_{-\mu^*}^{0}+\int_{-\mu^*}^{0}\frac{z e^z n\mu^{*^{n-1}}}{e^{z}+1}dz+
\end{align*}
\begin{align}
\int_{-\mu^*}^{0}\frac{z^2 e^z \frac{n\left(n-1\right)}{2}\mu^{*^{n-2}}}{e^{z}+1}dz+ ... \bigg]
\end{align}

Finally, from eqns. (10) and (13) we get the expression:

\begin{align}
\begin{split}
F_n\left(\mu^*\right)=\frac{{\mu^*}^{n+1}}{n+1}+{\mu^*}^nln\left(1+e^{-{\mu^*}}\right)-{\mu^*}^nln2\\+n{\mu^*}^{n-1}\left[{\mu^*} ln\left(1+e^{-{\mu^*}}\right)+Li_2\left(- e^{-{\mu^*}}\right)-\frac{\pi^2}{2}\right] + ...
\end{split}
\end{align}

That leaves the second part of the  integral in (9) namely 

$\int_{\mu^*}^{\infty}\frac{ x^n}{e^{x-\mu^*}+1}dx$ which will be calculated in detail in the appendix and will be shown to be small.

%                                                                                   thermal cond. wrt chemical pot
%Howdy be brave stay wild universe! $2^2 = \sqrt{16}$ \(x^2 + y^2 = z^2\) \[ x^n + y^n = z^n \]

\section{Extrema of the thermal conductivity} \label{TC-extema}

Minimizing electronic thermal conductivity is one of the possible pathways to increasing the figure of merit of the thermoelectric material. This task is achieved  in two steps in subsequent sections - 3.1 and 3.2 for its dependence on $\mu^*$ and then on $T$.

\subsection{Extrema of the thermal conductivity with respect to reduced chemical potential}

Following  Murali et al. \cite{murali2011} and taking the first derivative of the expression for $\sigma $ as a function of $\mu^*$ gives the following condition for the extremum, 
\begin{align*}
    \left[\Gamma \left( r+4 \right)-2\Gamma \left( r+3 \right) \right]Li_{r+1}^2Li_{r+2}-
\end{align*}
\begin{align}
\left(r+2\right)\Gamma \left( r+3 \right)Li_{r}Li_{r+2}^2=0
\end{align}
Rearranging gives:
\begin{align}
Li_{r+2}\Gamma \left( r+3 \right)\left\{\left[ \left(r+3\right)-2\right]Li_{r+1}^2- \left(r+2\right)Li_{r}Li_{r+2}\right\}=0
\end{align}and we get the two conditions, \( Li_{r+2}=0\) and \( \left(r+1\right)Li_{r+1}^2- \left(r+2\right)Li_{r}Li_{r+2}\)=0

Using series expansion for the polylogarithmic function gives for $r \geq 0$,
\begin{align}
Li_{r} \left(z\right)= \sum_{k=1}^{+\infty} \frac{z^k}{k^r}=z+\frac{z^2}{2^r}+\frac{z^3}{3^r} + ...
\end{align}

Case I : 
\(\mid z \mid\) \textless 1
\vskip .1in

For the approximation \(\mid z \mid\) \textless 1 which holds when $\mu^* < 0$, up to the order of $z^2$, one gets: 

\begin{align*}
     \left(r+1\right) \left(z+   \frac{z^2}{2^{r+1}}\right)^2 -
\end{align*}

\begin{align}
\left(r+2\right)\left(z+   \frac{z^2}{2^r}\right)\left(z+   \frac{z^2}{2^{r+2}}\right)=0
\end{align}
expanding and keeping terms up to $z^2$  one gets:

\begin{align}
\begin{split}
\left(r+1\right) \left(1+ \frac{z}{2^r} + \frac{z^2}{2^{2r+2}}\right)-
\left(r+2\right) \\ \left(1+\frac{z}{2^r} +\frac{z}{2^{r+2}}+\frac{z^2}{2^{2r+2}}\right)=0
\end{split}
\end{align}

Further simplification yields,

\begin{align}
\frac{z^2}{2^{r+2}}+\frac{z}{2^r}+\frac{rz}{2^{r+2}}+\frac{z}{2^{r+2}}=-1    
\end{align}
\begin{align}
\frac{z^2}{2^{r+2}}+\frac{z}{2^r}\left(1+\frac{r}{4}+\frac{1}{2}\right)=-1
\end{align}
\begin{align}
z\left[\frac{z}{4}+\left(\frac{3}{2}+\frac{r}{4}\right)\right]=-2^r
\end{align}

\begin{align}
e^{\mu^*}\left[-e^{\mu^*}+{r+6} \right]=2^{r+2}
\end{align}

Case II :
\(\mid z \mid\) \textless \textless 1

\begin{align}
e^{\mu^*}\left[-1-\mu^*+{r+6} \right]=2^{r+2}
\end{align}
\begin{align}
e^{\mu^*-(r+5)}\left[\mu^*-(r+5)  \right]=-2^{r+2}e^{-(r+5)}
\end{align}
\begin{align}
\mu^*-(r+5)=W\left(-2^{r+2}e^{-(r+5)}\right)
\end{align}

\begin{figure}[h!]
\centering
\includegraphics[scale=.6]{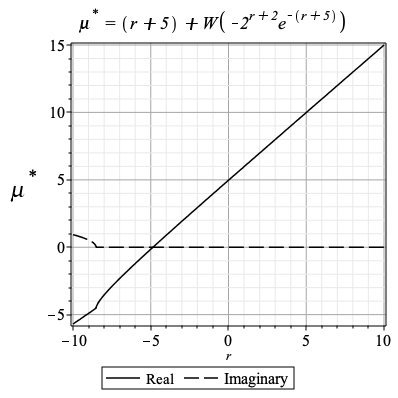}
\caption{Comparison of real and imaginary contributions to the $\mu^*$ in equation (26).}
\label{fig:planck_function}
\end{figure}

The expression for $\mu^*$ is illustrated in Fig. 2 below. In this figure we plot the real and imaginary parts of the expression. The imaginary part is zero above the value of $r$ corresponding the the beginning of the principal branch of the {LambertW} function. For the argument of W above, that value is $r =\frac{4-2 ln 2}{ln 2 -1} \approx -8.52$. It is of interest to note that for negative values of $r$ (e.g. $r = -3/2$ and $r=-1$), the Fermi-Dirac integrals have been tabulated by Blakemore \cite{blakemore1987}.

%                                                                 4     thermal conductivity wrt temperature

\subsection{Extrema of the thermal conductivity with respect to temperature}

We have the expression for thermal conductivity \(\lambda\)  in the form of polylogarithmic functions

\begin{align}
\begin{split}
\lambda_e =\frac {16\pi ml_0k (kT)^{r+2}}{3h^3}\bigg\{-\Gamma \left(r+4\right)Li_{r+3} \left(-e^{\mu^*}\right)\\+ \left(r+2\right)\Gamma \left(r+3\right)\frac{\left[Li_{r+2}\left(-e^{\mu^*}\right)\right]^2}{{Li_{r+1}\left(-e^{\mu^*}\right)}}\bigg\}
\end{split}
\end{align}

Differentiating and simplifying the above equation yields

\begin{align}
\begin{split}
\frac{d\lambda}{dT}=(r+2)\bigg\{-\left(r+3\right)Li_{r+3}\left(z\right)+ \left(r+2\right) \\ \frac{Li_{r+2}^2\left(z\right)}{Li_{r+2}\left(z\right)}\bigg\}-\mu^*Li_{r+2}\left(z\right)\bigg\{-\left(r+3\right)+\left(r+2\right) \\ \left[2-\frac{Li_r\left(z\right)Li_{r+1}\left(z\right)}{Li_{r+1}^2\left(z\right)}\right]\bigg\}=0
\end{split}
\end{align}

\begin{align}
Li_{r} \left(z\right)= \sum_{k=1}^{+\infty} \frac{z^k}{k^r}
\end{align}

Using the approximation  \(\mid z \mid\) \textless 1 , valid for $\mu^* < 0 $ and simplifying we get terms to $O(z^2)$:

\begin{align}
\begin{split}
\frac{z^2}{2^{2r+2}}\bigg[ -(r+2+\mu^*)+(r+2)\left(r+2-\mu^*\right)\frac{1}{4} \\ -\left(r+2\right)\left(r+3\right)\frac{1}{2}-\mu^*  \bigg] +\frac{z}{2^r}\bigg[-\left(r+2+\mu^*\right )  \\ -\left(r+2\right)\left(r+3\right)\frac{1}{8}  -\frac{\mu^*}{4} \bigg] = \left(r+2+\mu^*\right)
\end{split}
\end{align}

Case I:  \(\mid z \mid\)  \textless 1
\vskip 0.1in

Expansion of $z$ in terms of $\mu^*$, and setting $ A= 2(r+2)(r+9)+13 $ , $B=r+2$, $C= (r+2)(r+5) $, $ D = (r+2)(r+12)+16 $ , we get 

\begin{align}\label{off-log}
\begin{split}
e^{\mu^*} \bigg[\mu^{*^2} + \left(\frac{A}{D}\right)\mu^* + \frac{C}{D}\bigg]  = - \left(B+\mu^*\right)\left(\frac{2^{2r+5}}{D}\right)
\end{split}
\end{align}

\begin{align}\label{off-log-1}
e^{\mu^*}\left[\frac{\left(\mu^*-t_1\right)\left(\mu^*-t_2\right)}{\left(\mu^*+B\right)}\right]=- \frac{2^{2 r +2}}{D}
\end{align}

where $t_1$ and $t_2$ are the factors of the quadratic expression in $\mu^*$. The solution to equation (\ref{off-log-1}) can be written as a generalized Lambert W function \cite{maignan, valluri-res-gate},

\begin{align}
    \mu^* = W(t_1,t_2;-B;\frac{-2^{2r+2}}{D})
\end{align}

\vskip 0.2in

Case II:  \(\mid z \mid\)  \textless\textless 1
If we keep only terms of order $\mu^*$ , equation (\ref{off-log}) reduces to:

\begin{align}
e^{\mu^*}\frac{ \left(\mu^*+\frac{C}{A}\right)}{ \left(\mu^*+B\right)}=\frac{-2^{2r+5}}{A}
\end{align}

The solution in this case can also be obtained by use of the the generalized Lambert W function as:

\begin{align}
    \mu^* = W(\frac{-C}{A} , -B ; \frac{-2^{2r+5}}{A})
\end{align}

%                                                                      5                Electrical conductivity wrt temperature

\section{Extrema of the electrical conductivity with respect to temperature} \label{EC-extrema}

We have the following expression for the electrical conductivity \(\sigma\)

\begin{subequations}\label{38}

\begin{equation}\label{38a}
\sigma=\frac{16\pi me^2l_0\left(kT\right)^{r+1}\left(r+1\right)F_r}{3h^3}
\end{equation}

\begin{equation}\label{38b}
r\Gamma  \left(r\right)= \Gamma\left(r+1\right)
\end{equation}

\end{subequations}

\begin{subequations}\label{39}

\begin{equation}\label{39a}
\sigma=-\frac{16\pi me^2l_0\left(kT\right)^{r+1}\left(r+1\right)\Gamma\left(r+1\right)Li_{r+1}\left(-e^{\mu^*}\right)}{3h^3}
\end{equation}
Setting ,
\begin{equation}\label{39b}
c=-16\pi me^2l_0K^{r+1}\left(r+1\right)\Gamma\left(r+1\right)
\end{equation}

\end{subequations}

and differentiation of $\sigma$ gives 

\begin{align}
\frac{d\sigma}{dT}=cT^r\left[-\mu^*Li_r \left(z\right)+ \left(r+1\right)Li_{r+1} \left(z\right)\right]=0
\end{align}

Thereby, 

\begin{align}
-\mu^* \sum_{k=1}^{+\infty} \frac{z^k}{k^r}+\left(r+1\right)  \sum_{k=1}^{+\infty} \frac{z^k}{k^{r+1}}=0
\end{align}

Case I: $|z| < 1 $
\vskip 0.2in

For terms to cubic order, \(\mid z \mid\) \textless  1
\begin{subequations}\label{46}
\begin{equation}\label{46a}
-\mu^*\left(z+\frac{z^2}{2^r}+\frac{z^3}{3^r}  \right)+  \left(r+1\right)\left(z+\frac{z^2}{2^{r+1}}+\frac{z^3}{3^{r+1}}  \right)=0
\end{equation}
\begin{equation}\label{46b}
\begin{split}
z[\frac{1}{2^r}\left(-\mu^*+\frac{r+1}{2}\right)+\frac{z}{3^r}\left(-\mu^*+\frac{r+1}{3}\right) \\ +\frac{z^2}{4^r}\left(-\mu^*+\frac{r+1}{4}\right)]=\mu^*-\left(r+1\right)
\end{split}
\end{equation}
\end{subequations}

\begin{align}
\begin{split}
e^{\mu^*}\bigg[\frac{1}{2^r}\left(-\mu^*+\frac{a}{2}\right)-\frac{e^{\mu^*}}{3^r}\left(-\mu^*+\frac{a}{3}\right) \\ +\frac{e^{2\mu^*}}{4^r}\left(-\mu^*+\frac{a}{4}\right)\bigg]=a-\mu^*
\end{split}
\end{align}
where $a=r+1$.
On using a series approximation for the exponential function,

\begin{align}
\begin{split}
e^{\mu^*}\bigg[\frac{1}{2^r}\left(-\mu^*+\frac{a}{2}\right)-\frac{\left(1+\mu^*\right)\left(-\mu^*+\frac{a}{3}\right)}{3^r}  \\ +\frac{\left(1+2\mu^*\right)\left(-\mu^*+\frac{a}{4}\right)}{4^r}\bigg]=a-\mu^*
\end{split}
\end{align}

\begin{subequations}\label{49}
\begin{equation}\label{49a}
e^{\mu^*}\left(A-\mu^*B+\mu^{*^2}C\right)=a-\mu^*
\end{equation}
where,

\begin{equation}\label{49b}
A= \frac{a}{2^{r+1}}-\frac{a}{3^{r+1}}+\frac{a}{4^{r+1}}
\end{equation}
\begin{equation}\label{49c}
B= \frac{1}{2^r}+\frac{\frac{a}{3}-1}{3^r}+\frac{1-\frac{a}{2}}{4^r}
\end{equation}
\begin{equation}\label{49d}
C= \frac{1}{3^r}-\frac{2}{4^r}
\end{equation}
\end{subequations}

Neglecting \(\mu^{*^2}\),

\begin{subequations}\label{50}
\begin{equation}\label{50a}
e^{\mu^*}\left(A-\mu^*B\right)=a-\mu^*
\end{equation}
\begin{equation}\label{50b}
e^{\mu^*}\left[\frac{\mu^*B-A}{\mu^*-\left(a\right)}\right]=1
\end{equation}
\begin{equation}\label{50c}
e^{\mu^*}\left[\frac{\mu^*-\frac{A}{B}}{\mu^*-\left(a\right)}\right]=\frac{1}{B}
\end{equation}
\begin{equation}\label{50d}
\mu^*=W\left(\frac{A}{B},a,\frac{1}{B}\right)
\end{equation}
\end{subequations}

If  \(\mu^{*^2}\) is not neglected one ends up with:

\begin{align}
e^{\mu^*}\left(A-\mu^*B+\mu^{*^2}C\right)= a -\mu^*
\end{align}

Let $t_1$ and $t_2$ be the roots of the quadratic equation

\begin{align}
e^{\mu^*}\left[\left(\mu^*-t_1\right)\left(\mu^*-t_2\right)\right]=-\frac{\mu^*- a}{C}
\end{align}
The solution can be obtained by use of the generalized offset logarithmic function

\begin{align}
\mu^*=W\left(t_1,t_2,r+1,-\frac{1}{C}\right)
\end{align}
where,

\begin{align}
t_1 = \frac{\frac{B}{C}+\sqrt{\frac{B^2-4AC}{C^2}}}{2}
\end{align}
\begin{align}
t_2 = \frac{\frac{B}{C}-\sqrt{\frac{B^2-4AC}{C^2}}}{2}
\end{align}
Case II: $|z| << 1$
\vskip .2in
Retaining terms up to quadratic order, we have for  \(\mid z \mid\) \textless \textless 1 , valid for $\mu^* < 0$, we get:
\begin{subequations} \label{42}
\begin{equation} \label{42a}
-\mu^*\left(z+\frac{z^2}{2^r}  \right)+  \left(r+1\right)\left(z+\frac{z^2}{2^{r+1}}  \right)=0
\end{equation}
which simplifies to
\begin{equation} \label{42b}
-\mu^*\left(1+\frac{z}{2^r} \right)+  \left(r+1\right)\left(1+\frac{z}{2^{r+1}}  \right)=0
\end{equation}
\end{subequations}

Since \( z=-e^{\mu^*}  \), further simplification gives:

\begin{subequations}\label{43}
\begin{equation}\label{43a}
\left(r+1-\mu^*\right)+\left[\mu^*-\frac{\left(r+1\right)}{2}\right]\frac{e^{\mu^*}}{2^r}=0
\end{equation}
\begin{equation}\label{43b}
e^{\mu^*}=2^{r+1}\left[\frac{\mu^*-\left(r+1\right)}{\mu^*-\frac{\left(r+1\right)}{2}}\right]
\end{equation}
\end{subequations}

Rearrangement of the above equation ( \ref{43b} ) enables a solution by using the Offset logarithmic function with $\mu^*$ given by:
\begin{align}
\mu^*=G_k\left(2^{r+1},r+1,\frac{r+1}{2}\right)
\end{align}where $G_k$ is the generalized LambertW function.

\section{Minimization of the lattice thermal conductivity} \label{lat-extrema}
Following the approach developed by Cahill \textit{et al.} \cite{PhysRevB.46.6131}, we develop the conditions for minimum lattice thermal conductivity by extremizing the integrand of the following expression:

\begin{align} \label{debye_theta}
\kappa_{l,min} = \left(\frac{\pi}{6}\right)^{1/3}kn^{2/3}\sum_i v_i\left(\frac{T}{\Theta_i}\right)^2\int_{0}^{\frac{T}{\Theta_i}}\frac{x^3e^x}{\left(e^x-1\right)^2}dx
\end{align}

\begin{figure}[h!]
\centering
\includegraphics[scale=.3]{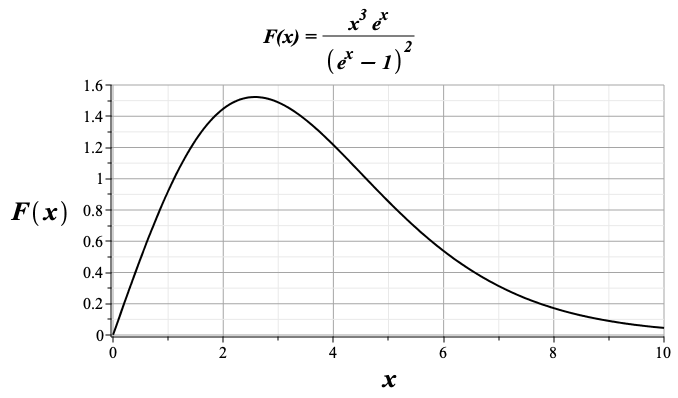}
\caption{Planck's blackbody radiation function.}
\label{fig:plank_function}
\end{figure}
\begin{align}\label{debye_integrand}
F\left(x\right)=\frac{x^3e^x}{{(e^x-1)}^2}
\end{align}\begin{align}
F^{'}\left(x\right)=\frac{{3x}^2e^x+x^3e^x}{{\left(e^x-1\right)}^2}+\frac{x^3e^x(-2)e^x}{{\left(e^x-1\right)}^3}=0 \notag
\end{align}
Further simplification results in the expression
\begin{align}\label{debye_soln}
e^x\frac{\left(3-x\right)}{\left(3+x\right)}=1
\end{align}

Numerical solutions of (\ref{debye_soln}) exist and the positive real value is 2.57568 which gives the maximum of equation (\ref{debye_integrand}). Exact solutions to the above equation can be obtained by use of the Offset Logarithm function.

We will now compute the integral in (\ref{debye_theta}) by setting $T/\theta_i = x$. The result is plotted in figure 4 multiplied by $(T/\theta_D)^2$ and shows a monotonically increasing function of $T/\theta_D$ with forbidden regions where real solutions do not exist. The plot exhibits forbidden regions. These interesting regions warrant further exploration that will be the subject of a separate study.

\begin{figure}[h!]\label{fig-kmin}
\centering
\includegraphics[scale=.5]{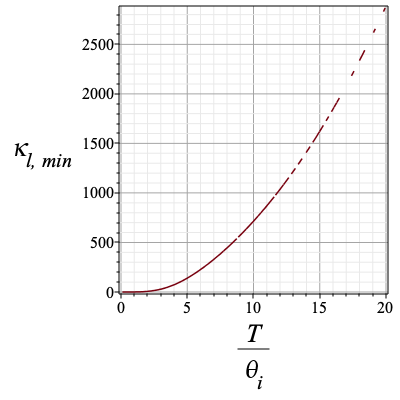}
\caption{$\kappa_{l.min}$ as a function $T/\theta_i$.}
\label{fig:plank_f2}
\end{figure}

The integral can be calculated by parts.

\begin{align}
\begin{split}
\int_0^x  {{\frac {{y}^{3}{{\rm e}^{y}}}{ \left( {{\rm e}^{y}}-1 \right) ^{2}}}}  =
3\,{\frac {{x}^{4}\ln  \left( -{{\rm e}^{x}}+1 \right) {{\rm e}^{x}}}{
{{\rm e}^{x}}-1}}-{\frac {{x}^{5}{{\rm e}^{x}}}{{{\rm e}^{x}}-1}} \\ +6\,{\frac {{x}^{3}{\it Li_2} \left({{\rm e}^{x}} \right) {{\rm e}^{x
}}}{{{\rm e}^{x}}-1}} -  3\,{\frac {{x}^{4}\ln  \left( -{{\rm e}^{x}} +1
 \right) }{{{\rm e}^{x}}-1}}+6\,{\frac {{x}^{2}\zeta \left( 3 \right) 
{{\rm e}^{x}}}{{{\rm e}^{x}}-1}} \\ -6\,{\frac {{x}^{2}{\it Li_3}
 \left({{\rm e}^{x}} \right) {{\rm e}^{x}}}{{{\rm e}^{x}}-1}}-6\,{
\frac {{x}^{3}{\it Li_2} \left({{\rm e}^{x}} \right) }{{{\rm e}^
{x}}-1}}  -6\,{\frac {{x}^{2}\zeta \left( 3 \right) }{{{\rm e}^{x}}-1}}+ \\
6\,{\frac {{x}^{2}{\it Li_3} \left({{\rm e}^{x}} \right) }{{
{\rm e}^{x}}-1}}
\end{split}
\end{align}

The preliminary analysis presented here shows that $\kappa_{min}$ is a monotonic function of $T/\theta_D$ as shown in figure 4. But further analysis is needed to determine the characteristics of the above expression representing $\kappa_{min}$ as a function of temperature (in units of $\theta_D$). It should be noted that for $x \neq 0 , \hspace{2mm} ln(1-e^x) < 0.$ giving complex values. 

\section{Conclusions} \label{conclusions}

In this work, a treatment using exact Fermi-Dirac integral expressions leads to the generalization of the Wiedemann-Franz  Law. The exact expression shows that non-negligible  corrections are possible. From figure 1, it is apparent that negative values of the index r of the polylog functions in the expression for the Lorenz number should be explored. The details of optimizing the figure of merit allowed us to explore the parameter space more carefully. Based on the results of optimizing the electronic thermal conductivity, we note that the chemical potential is strongly dependent on the polylog index $r$. Here again large negative values are possible. We also note that the imaginary part of the chemical potential is small compared to the real part and drops down to zero exactly at $r=-8.52$, which occurs at the start of the principal branch of the Lambert W function.

It is worth mentioning that the works of Poudel \textit{et al.} \cite{poudel-2008} and Zhao \textit{et al.} \cite{zhao_2014} have observed that a $ZT$ enhancement comes mainly from a reduction of the lattice thermal conductivity $\kappa_{min}$ and anomalously high Grüneisen parameters. \cite{zhao_2014}

Minimizing thermal conductivity seems to be the most effective path to enhancing the figure of merit. Thermal conductivity has both lattice and electronic contributions. In this work, both contributions have been carefully analyzed and minimization shows important restrictions on the complex solutions produced. The imaginary contributions have been found to be small and drop to negligible value as the argument of the Lambert W function function in the solutions approaches the value of $-1/e$.

Calculation of the extrema of the electrical and thermal conductivities leads to characteristic equations which give further insight into the ranges of r and $\mu*$ that favor an optimized figure of merit. The results of this paper should therefore be useful to experimental work directed at the chemistry and physics of materials chemistry looking into enhancing thermoelectric efficiency.

The stability and transport coefficients of the Lorenz number across different metals pose  a challenge to attaining a high $ZT$ value. Ouerdane et al.  have emphasized the importance of the study of electronic systems that undergo phase transitions and the role of fluctuating Cooper pairs.\cite{Ouerdane2015}
Further analytic work will concentrate on analyzing the properties of these characteristic equations and relating them to material properties. This combined analytical and numerical work will be the subject of a separate study.

\begin{acknowledgments}
Pranawa Deshmukh acknowledges the Shastri Indo Canadian Institute (SICI) for the award of an Indo-Canada collaborative research grant for this project. 
S.R. Valluri acknowledges the Natural Sciences Engineering Research Council (NSERC)
for a Discovery Grant during the course of which this work was performed. He would also like to thank Dr. K. N. Satyanarayana, the director of IIT Tirupati, for the gracious hospitality during his visit to the Institute where part of this research was performed.

\end{acknowledgments}

\section{Appendixes}

\appendix{Appendix I}\label{append1}
\vskip 0.2in
To compute the last term in eqn. (9) in section (2), let us call the integral represented in that term $\emph{I}$, where:

\begin{align}
\begin{split}
I = \int_{\mu^*}^{\infty}\frac{x^n}{e^{x-\mu^*}+1}dx = \int_{\mu^*}^{\infty}\frac{x^{n}e^{-(x-\mu^*)}}{1+e^{-(x-\mu^*)}}dx  
\\ 
= \int_{\mu^*}^{\infty}x^{n}e^{-(x-\mu^*)}\{1+e^{-(x-\mu^*)}\}^{-1}dx
\end{split}
\end{align}

A binomial expansion allows $\emph{I}$ to be written as:

\begin{align}
\begin{split}
I=\int_{\mu^*}^{\infty}x^{n}e^{-(x-\mu^*)}\bigg\{1+-e^{-(x-\mu^*)}+e^{-2(x-\mu^*)} \\ -  e^{-3(x-\mu^*)}  + e^{-4(x-\mu^*)}-e^{-5(x-\mu^*)}+ ... \\ + (-1)^{n}e^{-n(x-\mu^*)}+... \bigg\} dx
\end{split}
\end{align}

\begin{align}
\begin{split}
=\int_{\mu^*}^{\infty}\bigg\{x^{n}e^{-(x-\mu^*)}-x^{n}e^{-2(x-\mu^*)}+x^{n}e^{-3(x-\mu^*)} \\ -x^{n}e^{-4(x-\mu^*)}+...\bigg\}dx
\end{split}
\end{align}
Let us now define $I_{n,m}$ and integrate by parts, 

\begin{align}
\begin{split}
I_{n,m}=\int_{\mu^*}^{\infty}\{x^{n}e^{-m(x-\mu^*)}\}dx= \frac{x^{n}e^{-m(x-\mu^*)}}{-m}\bigg\rvert_{\mu^*}^{\infty} \\ +\frac{n}{m}\int_{\mu^*}^{\infty}\{x^{n-1}e^{-m(x-\mu^*)}\}dx
\end{split}
\end{align}

Where $m=1,2,3,.....$

Now, the definite integral $I_{n,m}$ can be written in the reduction form: 

\begin{align}
I_{n,m}= \frac{1}{m}(\mu^*)^{n}+\frac{n}{m}I_{n-1, m}
\end{align}
where, 
\begin{align}
\begin{split}
I_{n-1,m}=\int_{\mu^*}^{\infty}\{x^{n-1}e^{-m(x-\mu^*)}\}dx = \\ \frac{-x^{n-1}e^{-m(x-\mu^*)}}{m} \bigg\rvert_{\mu^*}^{\infty} +\frac{(n-1)I_{n-2,m}}{m}
\end{split}
\end{align}
Therefore,

\begin{align}
I_{n,m}=\frac{1}{m}(\mu^*)^{n}+\frac{n}{m}\{\frac{1}{m}(\mu^*)^{n-1}+\frac{(n-1)}{m}I_{n-2,m}\}
\end{align}

\begin{align}
=\frac{{\mu^*}^{n}}{m}+\frac{n}{m^2}(\mu^*)^{n-1}+\frac{n(n-1)}{m^2}I_{n-2,m}
\end{align}

\begin{align}
\begin{split}
= \frac{{\mu^*}^{n}}{m} \bigg\{1+\frac{n}{m}\frac{1}{\mu}+\frac{n(n-1)}{m^2}\frac{1}{{\mu^*}^2}+ \\ \frac{n(n-1)(n-2)}{m^3}\frac{1}{{\mu^*}^3}+...\bigg\}
\end{split}
\end{align}

\begin{align}
\begin{split}
   I_{n,1}=(\mu^*)^{n}\bigg\{1+\frac{n}{\mu^*} + \\ \frac{n(n-1)}{{\mu^*}^2}+\frac{n(n-1)(n-2)}{{\mu^*}^3}+...\bigg\} 
\end{split}
\end{align}

\begin{align}
\begin{split}
    I_{n,2}=\frac{(\mu^*)^{n}}{2}\bigg\{1+\frac{n}{2\mu^*}+\frac{n(n-1)}{4}\frac{1}{{\mu^*}^2} + \\ \frac{n(n-1)(n-2)}{8}\frac{1}{{\mu^*}^3}+...\bigg\}
\end{split}
\end{align}

\begin{align}
\begin{split}
    I_{n,3}=\frac{(\mu^*)^{n}}{3}\bigg\{1+\frac{n}{3\mu^*} + \frac{n(n-1)}{9}\frac{1}{{\mu^*}^2} + \\ \frac{n(n-1)(n-2)}{27}\frac{1}{{\mu^*}^3}+...\bigg\}
\end{split}
\end{align}

As $m$ becomes larger, $I_{n,m}$ decreases for a given $n$. the expression for the integral $(7)$ becomes:
\begin{align}
I=\int_{\mu^*}^{\infty}\frac{ x^n}{1+e^{x-\mu^*}}dx = I_{n,1} - I_{n,2}+I_{n,3}-I_{n,4}+...
\end{align}

Where
\begin{align}
\begin{split}
I_{n,1}= (\mu^*)^{n}\bigg\{1+ \frac{n}{\mu^*} +\frac{n(n-1)}{{\mu^*}^2} \\ + \frac{n(n-1)(n-2)}{{\mu^*}^3} + ... \bigg\}
\end{split}
\end{align}
and 
\begin{align}
I_{m,n} = \frac{1}{m}(\mu^*)^{n} + \frac{n}{m}I_{n-1,m}     
\end{align}
where $m= 2,3,4,...$ are given in equation 3. 

The expressions in $(3)$ are substituted in equation (14) to derive the expression for $F_{n}(\mu^*)$

Let 
\begin{align*}
\begin{split}
x - \mu^* = z \Rightarrow z = - \mu^* \ \text{when} \  x=0 \\
x = \mu^* + z  \Rightarrow z =0 \  \text{when} \  x = \mu^* \\
x^n = (\mu^*)^n (1 + \frac{z}{\mu^*})^n
\end{split}
\end{align*}

Using the transformation $x-\mu^*=z$ , the second term in equation (9) can be rewritten as:

\begin{align}
-\int_{0}^{\mu^*}\frac{x^{n}e^{x-\mu^x}}{1+e^{x-\mu^x}}dx= -\int_{-\mu^x}^{0}e^z\frac{\mu^{*n}}{1+e^z}(1+\frac{z}{\mu^*})^{n}dz
\end{align}

Using the Binomial expansion for $|z|<< 1$ and keeping terms up to second order in $z$, we have:

\begin{align}
=-\int_{-\mu^*}^{0} (\mu^*)^{n}\frac{e^z}{1+e^z}\{1+\frac{nz}{\mu^*}+\frac{n(n-1)}{2}\frac{z^2}{\mu^{*2}}+...\}dz
\end{align}

We can now integrate to get:

\begin{align*}
= -\Bigg[(\mu^*)^{n} \ln{(1+e^z)}\rvert_{-\mu^*}^{0}+n\int_{-\mu^*}^{0}z(\mu^*)^{n-1}\frac{e^z}{1+e^z}dz 
\end{align*}
\begin{align}+  \frac{n(n-1)}{2}\int_{-\mu^*}^{0}\frac{z^{2}e^{z}(\mu^*)^{n-2}}{1+e^z}dz+...\Bigg]
\end{align}
\begin{align}
\begin{split}
= -\Bigg[(\mu^*)^{n}\ln{(2)}-(\mu^*)^{n}\ln{(1+e^{-\mu^*})} + (\mu^*)^{n-1}n \\ \{\ln{(1+e^z)}z\rvert_{-\mu^*}^{0} 
-\int_{-\mu^*}^{0}\ln{(1+e^z)}dz \} + ... \Bigg] 
\end{split}
\end{align}
\begin{align}
\begin{split}
= - \Bigg[(\mu^*)^{n}\ln{(2)}-(\mu^*)^{n}\ln{(1+e^{-\mu^*})}  +n(\mu^*)^{n} \\ \ln{(1+e^{-\mu^*})}  - (\mu^*)^{n}n\int_{-\mu^*}^{0}\ln{(1+e^z)}dz + ... \Bigg]
\end{split}
\end{align}

For $e^{z}<1$ and $-\mu^{*}\leq z \leq 0$, and keeping terms up to $O(z)$ accuracy, we expand in powers of $z$:

\begin{align}
\begin{split}
=-\Bigg[(\mu^*)^{n}\{\ln{2}-\ln{(1+e^{-\mu^*})}\}  +n\{(\mu^*)^{n}  \ln{(1+e^{-\mu^*})} \\ -(\mu^*)^{n}\int_{-\mu^*}^{0}(e^z-\frac{e^{2z}}{2}+\frac{e^{3z}}{3}+...\} dz\Bigg]
\end{split}
\end{align}

\begin{align*}
=-\Bigg[(\mu^*)^{n}\{\ln{2}+(n-1)\ln{(1+e^{-\mu^*})}\} 
\end{align*}
\begin{align}
-\mu^{*n}(1-\frac{e^{-\mu^*}}{1^2}-\frac{1}{4}+\frac{e^{-2\mu^*}}{4}+\frac{1}{9}-\frac{e^{-3\mu^*}}{9}-...)\Bigg]
\end{align}

\begin{align}
\begin{split}
=-{\mu^*}^{n}\Bigg[\ln{2}+(n-1)\ln{(1+e^{-\mu^*})}-\{\frac{1}{1^2}-\frac{1}{2^2} \\ +\frac{1}{3^2}-\frac{1}{4^2}+...\}  +
\{\frac{e^{-\mu^*}}{1^2}-\frac{e^{-2\mu^*}}{2^2}+\frac{e^{-3\mu^*}}{3^2}+...\}\Bigg]
\end{split}
\end{align}

\appendix{Appendix II}\label{append2}
\vskip 0.2in
The term involving $z^2$ is given in the form:

\begin{align}
\begin{split}
\int_{-\mu^*}^0 \frac{(z^2 e^z)} {(e^z + 1)} dz = \bigg[{\mu^*}^2 (-\ln(e^{-\mu^*} + 1)) + \\
2 \mu^* Li_2(-e^{-{\mu^*}}) + 2 Li_3({-e^{-\mu^*}}) + \frac{3 \zeta(3)}{2}\bigg]
\end{split}
\end{align}
  for $e^{\mu^*}>=0$.
 \vskip 0.2in
This result can be obtained by expanding the integrand in powers of $e^z << 1$ , 
\begin{align}
-\bigg[\frac{n(n-1)(\mu^*)^{n-2}}{2}\int_{-\mu^*}^{0}\frac{z^{2}e^z}{1+e^z}dz \bigg]
\end{align}
Consider just the integral in the above expression. Integration by parts and then expanding $\ln(1+e^z)$ in powers of $e^z << 1$, we get: 
\begin{align}
\begin{split}
\int_{-\mu^*}^{0} \frac{z^{2} e^{z}}{1+e^z} dz =  z^{2}\ln{(1+e^z)}\rvert_{-\mu^*}^{0} - \\ 2\int_{-\mu^*}^{0}z\ln{(1+e^z)}dz
\end{split}
\end{align}

\begin{align}\begin{split}
=-(\mu^*)^{2}\ln{(1+e^{-\mu^*})} - 2\int_{-\mu^*}^{0}z\bigg\{e^{z} -\frac{e^{2z}}{2}+ \\ \frac{e^{3z}}{3}-\frac{e^{4z}}{4}+...\bigg\}dz
\end{split}
\end{align}

\begin{align}
\begin{split}
=-(\mu^*)^{2}\ln{(1+e^{-\mu^*})}-2\bigg[(\frac{e^z}{1}-  \frac{e^{2z}}{4}+  \frac{e^{3z}}{9}-  \\ \frac{e^{4z}}{16}+...)z\bigg\rvert_{-\mu^*}^{0}
  -\int_{-\mu^*}^{0}(\frac{e^z}{1}-\frac{e^{2z}}{4}+\frac{e^{3z}}{9}-\frac{e^{4z}}{16}+...)\bigg]
\end{split} 
\end{align}

Equation $(12)$ becomes:
\begin{align}
\begin{split}
-\frac{n(n-1)(\mu^*)^{n-2}}{2}\bigg[-(\mu^*)^{2}\ln{(1+e^{-\mu^*})}+ \\ 2\mu^{*}\{\frac{e^{-\mu^*}}{1^2}-\frac{e^{-2\mu^*}}{2^2}+\frac{e^{-3\mu^*}}{3^2}-\frac{e^{-4\mu^*}}{4^2}\} + \\ 2\{(\frac{e^z}{1}-\frac{e^{2z}}{8}+\frac{e^{3z}}{27}-\frac{e^{4z}}{64}+...)\rvert_{-\mu^*}^{0}\}\bigg]
\end{split}
\end{align}

Considering further now the second term $z^2$ in $\{\}$ of the $(1+\frac{z}{\mu^*})^n$ series.
Therefore:

\begin{align}
\begin{split}
-\frac{n(n-1)}{2}\bigg[(\mu^*)^{n}\ln{(1+e^{-\mu^*})}  +2(\mu^*)^{n-1} \\ \bigg\{\frac{e^{-\mu^*}}{1^2}-\frac{e^{-2\mu^*}}{2^2}+\frac{e^{-3\mu^*}}{3^2}-\frac{e^{-4\mu^*}}{4^2}+...\bigg\} \\
+2(\mu^*)^{n-2}\bigg\{\frac{1}{1^3}  -\frac{1}{2^3}+\frac{1}{3^3}-\frac{1}{4^3}+...-\frac{e^{-\mu^*}}{1^3} \\ +\frac{e^{-2\mu^*}}{2^3} -\frac{e^{-3\mu^*}}{3^3}+\frac{e^{-4\mu^*}}{4^3}-...\bigg\}\bigg]
\end{split}
\end{align}

\begin{align}
\begin{split}
=-\frac{n(n-1)}{2}\bigg[(\mu^*)^{n}\ln{(1+e^{-\mu^*})} +  \\ 2(\mu^*)^{n-1}(-\ln{(-e^{-\mu^*})} \\ 
+2(\mu^*)^{n-2}\bigg\{\sum_{l=1}^{\infty} \{\frac{(-1)^{l-1}}{l^3}-(-1)^{l-1}\frac{e^{l\mu^*}}{l^3}\}\bigg\}\bigg]
\end{split}
\end{align}

The next term in the $(1+\frac{z}{\mu^*})^n$ series is
\begin{align}
(\mu^*)^{n-3}\int_{-\mu^*}^{0}\frac{z^{3}\frac{(n(n-1)(n-2))e^{z}}{31}}{1+e^z}dz
\end{align}

\begin{align}
\begin{split}
= (\mu^*)^{n-3}\frac{n(n-1)(n-2)}{31}\bigg[z^3\ln{(1+e^z)}\rvert_{-\mu^*}^{0} \\ -3\int_{-\mu^*}^{0}\frac{z^{2}e^z}{e^{z}+1}dz\bigg]
\end{split}
\end{align}

\begin{align}
\begin{split}
=(\mu^*)^{n-3}n(n-1)[-(\mu^*)^{3}\ln{(1+e^{-\mu^*})}\frac{(n-2)}{31} \\ -\frac{(n-3)}{31}\int_{-\mu^*}^{0}\frac{z^{2}e^z}{1+e^z}dz]
\end{split}
\end{align}

\vskip 0.2in


\bibliography{references}% Produces the bibliography via BibTeX.

\end{document}